\documentclass[10pt,paper]{JHEP3}
\usepackage{graphics}
\usepackage{amsmath}

\makeatletter \@addtoreset{equation}{section} \makeatother

\def\unit{\hbox to 3.3pt{\hskip1.3pt \vrule height 7pt width .4pt \hskip.7pt
\vrule height 7.85pt width .4pt \kern-2.4pt \hrulefill \kern-3pt
\raise 4pt\hbox{\char'40}}}

\def\half{{\textstyle {1 \over 2}}}

\def\ap#1{\alpha^{\prime\,#1}}

\def\makeatletter{\catcode`\@=11}
\makeatletter
\def\mathbox#1{\hbox{$\m@th#1$}}%
\def\math@ccstyles#1#2#3#4#5#6#7{{\leavevmode
      \setbox0\mathbox{#6#7}%
      \setbox2\mathbox{#4#5}%
      \dimen@ #3%
      \baselineskip\z@\lineskiplimit#1\lineskip\z@
      \vbox{\ialign{##\crcr
             \hfil \kern #2\box2 \hfil\crcr
             \noalign{\kern\dimen@}%
             \hfil\box0\hfil\crcr}}}}
\def\mathaccstyles{\math@ccstyles\maxdimen}
\def\maththroughstyles{\math@ccstyles{-\maxdimen}}
\def\unitmatrixDT%
 {\maththroughstyles{.45\ht0}\z@\displaystyle {\mathchar"006C}\displaystyle 1}

\newcommand{\be}{\begin{equation}}
\newcommand{\ee}{\end{equation}}

\newcommand{\bea}{\begin{eqnarray}}
\newcommand{\eea}{\end{eqnarray}}
\newcommand{\nn}{\nonumber}

\newcommand{\beann}{\begin{eqnarray*}}
\newcommand{\eeann}{\end{eqnarray*}}

\newcommand{\dd}{{\rm d}}       
\newcommand{\pd}{\partial}      
\newcommand{\ve}{\varepsilon}   
\renewcommand{\L}{\mathcal{L}}  
\newcommand{\G}{\Gamma}         
\newcommand{\tr}{{\rm tr}}      
\newcommand{\Tr}{{\rm Tr}}      

\renewcommand{\a}{\alpha}

\newcommand{\g}{\gamma}
\renewcommand{\d}{\delta}
\newcommand{\e}{\epsilon}

\newcommand{\z}{\zeta}

\title{The effective action for the 4-point functions
  in abelian open superstring theory}
\author{Mees de Roo and Martijn~G.C.~Eenink\\
Institute for Theoretical Physics\\
   Nijenborgh 4, 9747 AG Groningen,\\
     The Netherlands\\
     E-mail: \email{m.de.roo@phys.rug.nl, m.g.c.eenink@phys.rug.nl }}

\preprint{UG-03/05\\ \hepth{0307211}}

\abstract{We construct the derivative corrections to the four-point
  vertices in the abelian open string effective action to
  all orders in $\ap{}$. The result
  is based on the structure of the string four-point
  function. Supersymmetry of these vertices is guaranteed by
  the supersymmetry of the $F^4$ term in the effective action.
  By this construction we establish the existence of an infinite
  number of supersymmetry invariants, the number of invariants
  at order $\ap{n}$ grows linearly with  $n$.}


\begin{document}

\section{Introduction\label{Intro}}

The problem of constructing the open superstring effective action
in ten dimensions is still not settled. Even in absence of
Chan-Paton factors (the abelian case) only a few sectors of the
complete effective action are known. The ten-dimensional
Born-Infeld action describes the dynamics for slowly varying
fields \cite{Fradkin}, which in the abelian case is a consistent
approximation. Its supersymmetric extension was derived in
\cite{CGNSW1,CGNSW2,ET,APS}. In \cite{AndTs} it was shown that
there are no corrections quadratic in derivatives to all orders in
$\ap{}$. All bosonic terms with four derivatives were derived in
\cite{Wyll}. Furthermore, it is known that there are no
corrections with an odd number of fields strengths\footnote{One
can show that as a consequence of the invariance of the theory
under worldsheet parity, all string amplitudes with an odd number
of external lines (involving only massless modes) vanish
\cite{Schwarz}. The authors were unaware of this whilst writing
\cite{CdRE03}, which led them to propose the above fact as a
conjecture. This footnote should settle the issue.}.

In this paper we derive a new all-order result. We  obtain the
effective action for the tree-level four-point function in the
abelian open superstring theory to all orders in $\ap{}$, i.e.,
including all derivative corrections. Our construction is an
example of the so-called S-matrix method \cite{Schwarz,GW,GS} to
construct the effective action\footnote{See \cite{MBM} for a
recent example of the use of the S-matrix method in the present
context.}. In this method one first writes down an action which
reproduces the propagators of the massless string modes, and
proceeds, in the absence of cubic interactions, to the four-point
function, which in string theory is non-polynomial in the momenta
$k_1,\ldots,k_4$ of the external particles. Because of the absence
of cubic interactions the four-point function does not have poles,
and the calculation of the four-point function only involves
one-particle-irreducible  diagrams.

One can easily write down a closed form for the effective action
because the open string four-point function factorizes in a
product of two terms: the first term ($K$) depending on
polarization vectors and wave functions, the second term (${\cal
G}$), proportional to the Veneziano amplitude,  depending only on
the momenta. The first term determines how the fields should
appear in the effective action. The second term expands into an
infinite series in $\ap{}$, and determines how derivatives should
be distributed over the fields. This structure applies to both the
bosonic terms and the terms involving fermions. Due to the
factorization of the amplitude, supersymmetry of the effective
action can be easily established. The supersymmetry of the
effective action which reproduces the term $K$ has been
established  a long time ago \cite{BRS}. The term ${\cal G}$, with
momenta replaced by derivatives acts on $K$ in the full effective
action, but we will show that the proof of supersymmetry still
works ``under the derivatives''.

In discussing the higher derivative contributions to the open
string effective action it is useful to introduce the following
notation \cite{CdRE03}. We write  such terms as
\begin{equation}
\label{notation}
   {\cal L}_{(m,n)} = {\ap{m}} \left(\partial^n F^{p}
        + \partial^{n+1}F^{p-2}\bar\chi\g\chi + \ldots\right)\,,
\end{equation}
For dimensional reasons we must have $2p-2m+n-4=0$.

The outline of this paper is as follows. In Section 2 we review
some properties of the tree-level four-point function in open
string theory, and construct the corresponding bosonic effective
action. We then proceed to discuss in Section 3 the fermionic
contributions and verify that the effective action is
supersymmetric. In Section 4 we consider the expansion of the
result in $\ap{}$, and give explicit results through order
$\ap{5}$. Conclusions are given in Section 5.


\section{The 4-photon tree amplitude and its effective action}

The open string tree-level 4-point function is given by
\cite{Schwarz}: \be \label{ampl}
  \mathcal{A}(1,2,3,4) =
  -16\,i\,g^2\ap{2}(2\pi)^{10}\d^{(10)}(k_1+k_2+k_3+k_4)\,
  \mathcal{G}(k_1,k_2,k_3,k_4)\,K(1,2,3,4)
\ee
$\mathcal{G}$ contains the $\ap{}$ dependence and is given by:
  \be \label{functionG}
    \begin{split}
    \mathcal{G}(k_1,k_2,k_3,k_4)
    &= G(s,t) + G(t,u) + G(u,s) \\
    &= \frac{\G(-\a's)\G(-\a't)}{\G(1-\a's-\a't)}+
       \frac{\G(-\a't)\G(-\a'u)}{\G(1-\a't-\a'u)}+
       \frac{\G(-\a'u)\G(-\a's)}{\G(1-\a'u-\a's)}.
    \end{split} \,.
  \ee
Here $s$, $t$, and $u$ are the Mandelstam variables, satisfying
$s+t+u=0$. They are defined in terms of the $k_i$ only up to
momentum conservation and the mass-shell condition. We choose to
write them in such a way that $\mathcal{G}$ is manifestly
symmetric in the $k_i$:
  \be \label{Mandelstam}
    \begin{split}
    s &= -\,k_1\cdot k_2-k_3\cdot k_4, \\
    t &= -\,k_1\cdot k_3-k_2\cdot k_4, \\
    u &= -\,k_1\cdot k_4-k_2\cdot k_3.
    \end{split}
  \ee
As discussed in the above, $\mathcal{G}$ is regular as
$k_i\rightarrow 0$, which one can verify by expanding
(\ref{functionG}) in $\ap{}$. For now we just mention that
\begin{equation}
    \mathcal{G}(k_1,k_2,k_3,k_4) = -\frac{\pi^2}{2}+\mathcal{O}(\ap{2}),
\end{equation}
and postpone a detailed discussion of the expansion to a later
section. $K$ involves not only the momenta of the external
particles, but also their wavefunctions. For the 4-boson amplitude
we have:
  \be \label{kinematicfactor}
    K(1,2,3,4) = t^{abcdefgh} k^1_{a}\z^1_{b}
    k^2_{c}\z^2_{d} k^3_{e}\z^3_{f} k^4_{g}\z^4_{h},
  \ee
where $\z^i$ is the polarization vector of the $i$th incoming
photon, and the tensor $t$ is defined in Appendix A. The leading
order contribution to the amplitude is just
(\ref{kinematicfactor}) times a constant. It is well known that it
is reproduced by the action
\begin{equation} \label{L(2,0)boson}
  \begin{split}
  S_{(2,0)}
  &= \frac{1}{8}(2\pi g\a')^2\int\dd^{10}x\,
  \left( \tr F^4 - \frac{1}{4}(\tr F^2)^2\right) \\
  &= \frac{1}{8}(2\pi g\a')^2\int\dd^{10}x\,
  \frac{1}{24}t_{abcdefgh}F^{ab}F^{cd}F^{ef}F^{gh}\,,
  \end{split}
\end{equation}
We observe that every factor of momentum $k_i$ in
(\ref{kinematicfactor}) is reproduced by a derivative acting on
the appropriate field in (\ref{L(2,0)boson}).

The complete amplitude (\ref{ampl}) differs from the leading order
contribution by multiplication with $\mathcal{G}$, i.e. by extra
factors of momentum. In order to reproduce these factors, we
simply need to act with derivatives on the appropriate fields.
This is implemented by first allowing the four fields to be
``defined at different points in spacetime", resulting in a
non-local action. That is, we consider the fields $A_a(x_i)$,
where $i=1,\ldots,4$, and then replace the momenta $k_i$ in the
amplitude by differentiations with respect to the appropriate
coordinate in the effective action, i.e. $k_{i,a} \rightarrow
-i\pd/\pd x_i^a$. We need to multiply the resulting expression by
delta functions and then integrate over the $x_i$ to make the
action local.

Hence we define the following differential operator
\begin{equation} \label{operatorD}
  D(\pd_{x_1},\pd_{x_2},\pd_{x_3},\pd_{x_4}) \equiv
  \left.\mathcal{G}(k_1,k_2,k_3,k_4)\right|_{k_i\rightarrow
  -i\pd_{x_i}},
\end{equation}
which we use to write down the effective action for the complete
four-photon amplitude:
\begin{multline} \label{effactboson}
  S_{{\rm eff}}[A_a] = -\frac{1}{24}(g\a')^2\int\dd^{10}x
  \left\{\prod_i\dd^{10}x_i\,\d^{(10)}(x-x_i)\right\}
  D(\pd_{x_1},\pd_{x_2},\pd_{x_3},\pd_{x_4})\\
  \times t_{abcdefgh}F^{ab}(x_1)F^{cd}(x_2)F^{ef}(x_3)F^{gh}(x_4).
\end{multline}
$D$ is understood as a Taylor expansion in $\ap{}$. Then the
multiple integral over the $x_i$ factorizes into a product of
integrals, each involving only one of the $x_i$ and none of the
others, which is necessary in order that the above expression is
well defined. The actual proof that this action reproduces the
amplitude (\ref{ampl}) can be found in Appendix B.

As mentioned above, we choose to express $s,t,u$ in terms of the
$k_i$ in such a way that $\mathcal{G}$ is manifestly symmetric in
the momenta. This will turn out to be convenient in the following
section. It is not difficult to see that a different prescription
than (\ref{Mandelstam}) would result in modifications of the
effective action (\ref{effactboson}) by total derivatives and/or
the effects of field redefinitions\footnote{Remember that terms
containing lowest order field equations can be induced in the
effective action by means of a redefinition of the fields. See
e.g. \cite{CdRE02,CdRE03}.}. This follows from momentum
conservation $k_1^a+k_2^a+k_3^a+k_4^a=0$ and the mass-shell
conditions $k_i^2=0$, respectively.


\section{The fermionic contributions and supersymmetry}

As is well known, the supersymmetric extension of
(\ref{L(2,0)boson}) is unique and given by \cite{BRS,goteborg}
\footnote{Supersymmetry holds only order by order in the number of
 fields,
 starting with the standard super-Maxwell action $F^2 +
\bar\chi\gamma\partial\chi$, and requires modifications of the
supersymmetry transformations at all orders. The
superinvariants involving higher-derivative terms defined below
have a similar structure.}:
\begin{equation}\label{L(2,0)}
  \begin{split}
  S_{(2,0)}= \frac{1}{8}(2\pi g\a')^2\int\dd^{10}x\,
  \bigg( & \tr F^4 - \frac{1}{4}(\tr F^2)^2 \\
  &
  -2F_{ab}F_{ac}\bar{\chi}\g_b\pd_c\chi +
  F_{ab}F_{cd}\bar{\chi}\g_{abc}\pd_d\chi +
  \frac{1}{3}\bar{\chi}\g_a\pd_b\chi\bar{\chi}\g_a\pd_b\chi \bigg).
  \end{split}
\end{equation}
This action reproduces the four-point string amplitudes involving
two and four fermions \cite{BBRS} to lowest order in $\ap{}$. It
is then easy to guess what the effective action should be when
fermionic interactions as well as higher derivative corrections
are included:
\begin{multline}\label{effact}
  S_{{\rm eff}}[A_a,\chi] = -(g\a')^2\int\dd^{10}x
  \left\{\prod_i\dd^{10}x_i\,\d^{(10)}(x-x_i)\right\}
  D(\pd_{x_1},\pd_{x_2},\pd_{x_3},\pd_{x_4})
  \\
  \begin{split}
  \times& \bigg\{ F_{ab}(x_1)F_{bc}(x_2)F_{cd}(x_3)F_{da}(x_4)
    - \frac{1}{4} F_{ab}(x_1)F_{ab}(x_2)F_{cd}(x_3)F_{cd}(x_4) \\
  &  - 2F_{ab}(x_1)F_{ac}(x_2)\bar{\chi}(x_3)\g_b\pd_c\chi(x_4) +
    F_{ab}(x_1)F_{cd}(x_2)\bar{\chi}(x_3)\g_{abc}\pd_d\chi(x_4) \\
  &
  +\frac{1}{3}\bar{\chi}(x_1)\g_a\pd_b\chi(x_2)\bar{\chi}(x_3)\g_a\pd_b\chi(x_4)\bigg\}
  \end{split}
\end{multline}
It is not difficult to prove that this action is supersymmetric.
As explained in the previous section, the operator $D$ is
symmetric in the $\pd_{x_i}$. This implies that, when we apply the
Noether method\footnote{For a detailed description of the Noether
method in the case of super Yang-Mills theory we refer to our
previous papers \cite{CdRE02,CdRE03} with A.~Collinucci.} 
to (\ref{effact}), we can
perform the same manipulations as the ones necessary to
demonstrate the supersymmetry of (\ref{L(2,0)}).\\
Consider for example the variation of the first term in
(\ref{L(2,0)}). It is given by
\begin{align}
  \d\,\big(\tr\,F^4\big) =\,& \d F_{ab}F_{bc}F_{cd}F_{da} +
  F_{ab}\d F_{bc}F_{cd}F_{da} + F_{ab}F_{bc}\d F_{cd}F_{da} + F_{ab}F_{bc}F_{cd}\d F_{da}\nn\\
  =\,& 4\, F_{ab}F_{bc}F_{cd}\d F_{da}.
\end{align}
The last step is of course completely trivial in the local case,
but essential for proving the supersymmetry. In the non-local case
(\ref{effact}), this last step is not automatic. We see
that it is the symmetry of $D$ that allows us to perform it.\\
In addition to algebraic manipulations of the kind described
above, it is also necessary to perform partial integrations to
prove the supersymmetry. In the local case one encounters for
example the following total derivative at an intermediate stage of
the calculation:
\begin{equation}
  \pd_a\,\big( F_{ab}\,\tr\,F^2\,\bar{\e}\g_b\chi\big).
\end{equation}
In the non-local case this term will manifest itself as
\begin{equation}
  \left( \frac{\pd}{\pd x_1^a}+\frac{\pd}{\pd x_2^a}+\frac{\pd}{\pd
  x_3^a}+\frac{\pd}{\pd x_4^a} \right)
  F_{ab}(x_1)F_{cd}(x_2)F_{cd}(x_3)\,\bar{\e}\g_b\chi(x_4).
\end{equation}
This still gives rise to a total derivative, since we can pull the
$\sum_i \pd/\pd x^a_i$ out of the integration over the $x_i$:
\begin{equation}
  \begin{split}
  &\int\dd^{10}x
  \left\{\prod_i\dd^{10}x_i\,\d^{(10)}(x-x_i)\right\}
  D(\pd_{x_1},\pd_{x_2},\pd_{x_3},\pd_{x_4}) \\
  &\qquad\qquad\times\left( \sum_j\frac{\pd}{\pd x^a_j}\right)
  F_{ab}(x_1)F_{cd}(x_2)F_{cd}(x_3)\,\bar{\e}\g_b\chi(x_4) \\
  &=
  \int\dd^{10}x\,\frac{\pd}{\pd x}
  \int\left\{\prod_i\dd^{10}x_i\,\d^{(10)}(x-x_i)\right\}
  D(\pd_{x_1},\pd_{x_2},\pd_{x_3},\pd_{x_4})\\
  &\qquad\qquad\times
  F_{ab}(x_1)F_{cd}(x_2)F_{cd}(x_3)\,\bar{\e}\g_b\chi(x_4).
  \end{split}
\end{equation}
Here the symmetry properties of $D$ are not required.\\
We conclude, that the fact that (\ref{effact}) is supersymmetric
follows immediately from the supersymmetry of (\ref{L(2,0)}).

The above actually shows that when we replace $D$ in
(\ref{effact}) by {\em any}\ symmetric differential operator
$\Delta(\pd_{x_1},\ldots,\pd_{x_4})$, we obtain a supersymmetric
action.


\section{Derivative expansion of the effective
action}\label{expansionsection}

In this section we will consider the derivative expansion of the
effective action (\ref{effactboson}). This will allow us to make
contact with previously obtained results at order $\a'^4$ as well
as to present new results at order $\a'^5$. But first let us
discuss the form of the generic Lorentz invariant symmetric
differential operator $\Delta(\pd_{x_1},\ldots,\pd_{x_4})$ and
determine the number of independent supersymmetric invariants that
are possible at any given order in $\ap{}$.

To find the form of $\Delta(\pd_{x_1},\ldots,\pd_{x_4})$ we need
the most general Lorentz invariant expression that is symmetric
and regular in the momenta $k_i$, after which we substitute
$k_i\rightarrow -i\pd_i$. In such an expression only combinations
$k_i\cdot k_j$ and their products can enter\footnote{We do not
have to consider contractions with the $\ve$-tensor, since all
scalars that one can form by contracting it with the momenta $k_i$
vanish.}. Using momentum conservation and the mass-shell condition
all such terms can be written as combinations of $s,t,u$. Any
completely symmetric polynomial in $s,t,u$ can be written as:
\begin{equation} \label{P(klm)polynomial}
  \sum_{k\leq l\leq m}\,\a'^{k+l+m}\,c_{k,l,m}\,\mathcal{P}(k,l,m),
\end{equation}
where the $c_{k,l,m}$ are constants and
\begin{equation}
  \mathcal{P}(k,l,m) =
  s^kt^lu^m + s^kt^mu^l + s^mt^ku^l + s^mt^lu^k + s^lt^mu^k +
  s^lt^ku^m.
\end{equation}
Define
\begin{equation}
  P(n) = s^n+t^n+u^n\,,\qquad Q = stu\,.
\end{equation}
$\mathcal{P}(k,l,m)$ can be expressed in terms of $P(n)$ and $Q$:
\begin{equation}
  \mathcal{P}(k,l,m)=Q^k\big(P(l-k)P(m-k)-P(l+m-2k)\big).
\end{equation}
Furthermore, it follows from $P(1)P(n-1)=0$ that
\begin{equation} \label{P-relation}
  P(n)=\tfrac{1}{2}P(2)P(n-2)+QP(n-3).
\end{equation}
We conclude that we can express (\ref{P(klm)polynomial}) in powers
of $P\equiv P(2)$ and $Q$:
\begin{equation}
  \sum_{a,b}\,\a'^{2a+3b}\,d_{a,b}\,P^aQ^b,
\end{equation}
where the $d_{a,b}$ are constants. The number $N_{P,Q}(m)$ of
possible independent combinations of $P$ and $Q$, at order $\a'^m$
in the above expansion, is given by
\begin{equation} \label{numberinvariants}
  N_{P,Q}(m) =
  \left\{
  \begin{array}{ll}
    \left[m/6\right]+1, & \quad\mbox{ if }\quad m\neq 6\times\left[m/6\right]+1\\
    \left[m/6\right], & \quad\mbox{ if }\quad m = 6\times\left[m/6\right]+1,
  \end{array}
  \right.
\end{equation}
where $[x]$ denotes the largest integer smaller than $x$.

This implies that, for a given $m$, there are $N_{P,Q}(m)$
independent supersymmetric
contributions to the open string tree-level effective action that
contain terms of the form $\pd^{2m}F^4$.

We now turn to the derivative expansion of (\ref{effact}). We use
the Taylor expansion for $\log \Gamma(1+z)$,
\begin{equation}
  \log \Gamma(1+z) = -\g z +
  \sum_{m=2}^{\infty}(-1)^m\zeta(m)\frac{z^m}{m},
\end{equation}
where $\zeta(n)$  is the Riemann zeta-function, $\g$ the
Euler-Mascheroni constant, to obtain the following expression for
$G(s,t)$:
\begin{equation}
   \ap{2}\,G(s,t) = \frac{1}{st}\exp\big\{\sum_{m=2}^{\infty}\ap{m}\frac{\zeta(m)}{m}
              (s^m+t^m-(s+t)^m)\big\}.
\end{equation}
This expression can be used to calculate the $\ap{}$ expansion of
$\mathcal{G}(k_1,\ldots,k_4)$. We give here the first terms in
this expansion, expressed in $P$ and $Q$:
\begin{align}
 \mathcal{G}(k_1,\ldots,k_4) =&
 -\half\,\pi^2
 -\tfrac{1}{48}\ap{2}\,\pi^4\,P
 -\tfrac{1}{2}\a'^3\pi^2\zeta(3)\,Q
 \\
 &-\tfrac{1}{960}\a'^4\,\pi^6\,P^2
  -\tfrac{1}{48} \a'^5\pi^2\,
      \big( \pi^2\zeta(3)+12\,\zeta(5)\big)PQ
 \nn\\
 & -\tfrac{1}{967680}\a'^6
     \Big(51\pi^8\,P^3 + 8\pi^2\big(31\pi^6 + 30240\,\zeta(3)^2\big)\,Q^2\Big)
 \nn\\
 & -\tfrac{1}{960}\a'^7\,
   \pi^2\big( \pi^4\zeta(3) + 10\,\pi^2\zeta(5) + 120\,\zeta(7)\big)\,P^2Q
 \nn\\
 & -\tfrac{1}{58060800}\a'^8
   \Big(155\pi^{10}\,P^4 + 32\pi^2
    \big(67\pi^8 + 18900\,\pi^2\zeta(3)^2+ 453600\,\zeta(3)\zeta(5)\big)P Q^2 \Big)
 \nn\\
 & -\tfrac{1}{967680}\a'^9
  \Big(\pi^2\big(51\pi^{6}\zeta(3) + 504\,\pi^4\zeta(5)
  + 5040\,\pi^2\zeta(7) + 60480\,\zeta(9)\big)P^3Q
 \nn\\
 &\qquad\qquad
  + 8\,\pi^2\big(31\pi^6\zeta(3) +
  10080\,(\zeta(3)^3+2\,\zeta(9))\big)Q^3
  \Big)+\ldots
 \nn
\end{align}
We see that, at least to this order, all possible combinations of
$P$ and $Q$ indeed appear. String theory thus seems to make use of
all available superinvariants. By substituting derivatives for
momenta in the above expansion and inserting the resulting
expression in (\ref{effact}), one can straightforwardly construct
the contribution to the effective action at any desired order in
$\ap{}$. We demonstrate this for the bosonic terms at order
$\a'^4$ and $\a'^5$. At order $\a'^4$ we obtain:
\begin{eqnarray}
   \L_{(4,4)}
   &=& \tfrac{1}{288}\pi^4\,g^2\a'^4\,
   t_{abcdefgh}\,\pd_kF_{ab}\pd_kF_{cd}\pd_lF_{ef}\pd_lF_{gh} \nn
   \\
   &=& \tfrac{1}{36}\pi^4\,g^2\a'^4
   \Big(\big(\pd_kF_{ab}\pd_lF_{bc}\pd_kF_{cd}\pd_lF_{da}
   +2\,\pd_kF_{ab}\pd_kF_{bc}\pd_lF_{cd}\pd_lF_{da}\big)
   \\
   &&
   \qquad\qquad\qquad -\tfrac{1}{4}\big(\pd_kF_{ab}\pd_kF_{ab}\pd_lF_{cd}\pd_lF_{cd}
   +2\,\pd_kF_{ab}\pd_lF_{ab}\pd_kF_{cd}\pd_lF_{cd}\big)\Big). \nn
\end{eqnarray}
This expression is consistent with results obtained previously by
different methods \cite{KS3}. We have also checked explicitly that
the terms bilinear in the fermions - which were obtained in
\cite{CdRE03}, see also \cite{DHHK} - are reproduced correctly. As
always, this result is determined up to total derivatives and
terms containing lowest order field equations.

At order $\a'^5$ we obtain the following result:
\begin{eqnarray}
\label{act56}
  \L_{(5,6)}
  &=& -\tfrac{1}{6}\pi^2\z(3)\,g^2\a'^5\,
  t_{abcdefgh}\,\pd_k\pd_l\pd_mF_{ab}\pd_kF_{cd}\pd_lF_{ef}\pd_mF_{gh} \\
  &=& -4\pi^2\z(3)\,g^2\a'^5
  \Big(
  \pd_k\pd_l\pd_mF_{ab}\pd_kF_{bc}\pd_lF_{cd}\pd_mF_{da}
  -\tfrac{1}{4}\,\pd_k\pd_l\pd_mF_{ab}\pd_kF_{ab}\pd_lF_{cd}\pd_mF_{cd}\Big).
  \nn
\end{eqnarray}

As was already mentioned above, each of the terms 
${\cal L}_{(m,2m-4)}$
constructed in this paper are, together with the order $\ap{0}$
super-Maxwell action, supersymmetric to fourth order in the 
number of fields. 
From the
point of view of the Noether procedure each of these terms
contributes to genuine superinvariants that extend to all orders 
in the number of fields. 
One such superinvariant is the complete open superstring
effective action, to which all $\L_{(m,2m-4)}$ contribute. 
One can then pose the question how many independent 
sub-invariants the string effective action contains. 
In \cite{CdRE03} the
general structure of the web of supersymmetric derivative
corrections was discussed in some detail.  
It was argued there that the contributions
which in the string effective action have coefficients involving
powers of $\zeta(n)$, $n$ odd, only,  should form independent 
invariants.\\ 
The simplest assumption, which was posed as a conjecture
in \cite{CdRE03}, is
that the sectors ${\L}_{(2,0)}$, ${\L}_{(4,4)}$ and
${\L}_{(m,2m-4)}$, $m$ odd, contain the next-to-leading-order
contributions to separate superinvariants, and that there are no
other all-order invariants starting at $\L_{(m,n)}$ for any $m,n$. 
The results obtained in the present paper do not falsify this
conjecture.\\
Note that the conjecture implies that the terms involving, for example,
$\zeta(9)P^3Q$ and $\zeta(9)Q^3$, which are independently invariant
when supersymmetry to fourth order in the number of fields 
is considered, should
become part of a single invariant if supersymmetry is required also at 
higher orders.


\section{Summary and conclusions}

We have obtained a new result to all orders in $\ap{}$ for a
specific sector of the open superstring effective action: the
four-point vertices. The bosonic four-derivative term agrees with
\cite{AndTs}, the fermionic contributions at that order agree with
our result \cite{CdRE03}, which was obtained with the Noether
procedure.

The bosonic part of the term at order $\ap{5}$ (\ref{act56}) (six
derivatives) can be compared with a conjecture by Wyllard
\cite{Wyll2}.

In \cite{Wyll2} it was conjectured that all derivative corrections
to the Born-Infeld action follow from the corrections to the
Wess-Zumino term. This conjecture is applied in \cite{Wyll2} using
the results for the Wess-Zumino term of \cite{Wyll} as input. We
have taken the six-derivative corrections given in formula (4.16)
of \cite{Wyll2}, and extracted the terms of fourth order in $F$.
We find:
\begin{equation}
  \L_{(5,6)\,{\rm Wyllard}} = -4\pi^2\z(3)\,g^2\a'^5
   \Big(
  \pd_k\pd_lF_{ab}\pd_k\pd_mF_{bc}\pd_l\pd_mF_{cd} F_{da}
  -\tfrac{1}{4}\,
  \pd_k\pd_lF_{ab}\pd_k\pd_mF_{ab}\pd_l\pd_mF_{cd}F_{cd}\Big)
\end{equation}
This agrees, up to field redefinitions, with our result
(\ref{act56}). However, this agreement should be interpreted with
care. First of all the procedure of \cite{Wyll2} involves an
infinite series involving functional derivatives of the
Born-Infeld action with respect to the field strength $F$. The
conjecture requires an ordering prescription for these functional
derivatives. For our comparison we have taken the simplest
solution to this ordering ambiguity. Secondly, the corrections to
the Wess-Zumino term in \cite{Wyll} are not complete. Other
corrections, such as those evaluated in
\cite{Mukhi1,Mukhi2,Grange}, will contribute as well. On applying
Wyllard's proposal to these extra terms, further six-derivative
corrections to the Born-Infeld term might be generated. Our
agreement with \cite{Wyll2} indicates that these extra terms do
not give rise to new six-derivative $F^4$ terms in the Born-Infeld
action\footnote{We are grateful to Niclas Wyllard for useful
remarks and suggestions on these issues.}.

\begin{figure} \label{expansionfigure}
  \begin{center}
    \scalebox{0.80}{\input{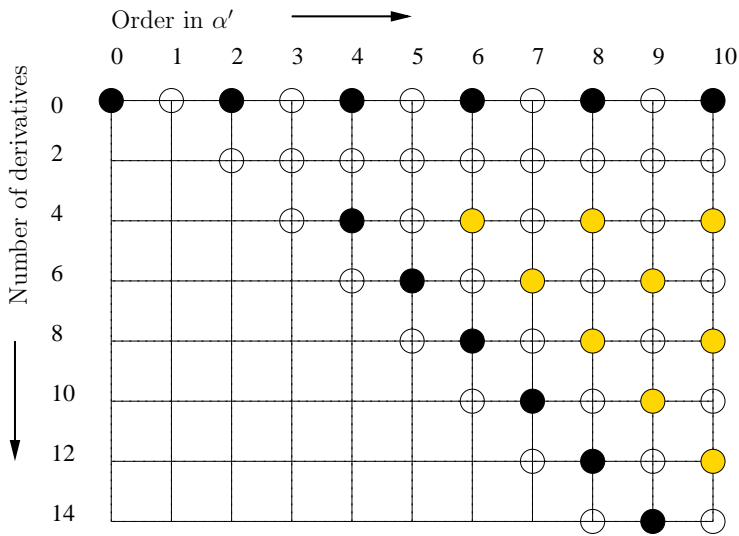}}
  \end{center}
  \caption{Structure of the abelian open superstring tree level effective
    action. Black dots indicate nonempty sectors of which the
    explicit form is known. Empty white dots correspond to sectors
    that are known to be empty up to field redefinitions. The yellow dots indicate
    sectors that are known to be nonempty but have yet to be constructed explicitly.}
\end{figure}

In figure 1 we show the present situation for the effective action
of abelian open superstring. Black dots indicate sectors for which
bosonic as well as fermionic terms are known, and supersymmetry
has been established. The terms corresponding to the four-point
function established in this paper are along the line $(m,2m-4)$,
where $m$ is the order of $\ap{}$. All bosonic four-derivative
terms have been given in \cite{Wyll}, but the fermionic
contributions remain to be found. Clearly further progress
requires a better understanding of the six- and higher-point
functions from string theory. In the case of the four-point
function supersymmetry of ${\cal L}_{(m,2m-4)}$ for $m>0$ follows
from the supersymmetry of ${\cal L}_{(2,0)}$. The generalization
one could hope for is that supersymmetry of the full effective
action follows, ``under the derivatives'', from supersymmetry of
the Born-Infeld action.

An interesting problem is the extension of our result to the
nonabelian case. In that case (\ref{ampl}) is still valid, but
$\mathcal{G}$ contains now also the group structure:
\be \label{functionGnab}
    \mathcal{G}(k_1,k_2,k_3,k_4)
    = (t_{ABCD}+t_{DCBA})G(s,t) +
       (t_{ABDC}+t_{CDBA})G(t,u) +
       (t_{ACBD}+t_{DBCA})G(u,s)\,,
\ee where $t_{ABCD}= \Tr\lambda_A\lambda_B\lambda_C\lambda_D$. The
problem is now that at order $\ap{n}$ we are not just discussing
the derivative correction to the four-point function, but also
contributions with different numbers of derivatives and $F$'s.
These all communicate through the relation $[{\cal D}, {\cal
D}]F=[F,F]$, and correspond to vertical lines in Figure 1. The
sectors which are independent in the abelian case are connected in
the nonabelian situation. The method of \cite{KS3} to organize the
nonabelian effective string action in terms of symmetric traces
seems to maximize the usefulness of the abelian results for
solving the nonabelian problem. Nevertheless, making further
progress with the nonabelian case remains a formidable problem.


\section*{Acknowledgements}

We are grateful to N.~Wyllard, S.~Mukhi and R.~Medina for useful
remarks. The work of Martijn Eenink is part of the research
programme of the ``Stichting voor Fundamenteel Onderzoek van de
Materie'' (FOM). This work is supported in part by the European
Commission RTN programme HPRN-CT-2000-00131, in which we are
associated to the University of Utrecht.

\appendix

\section{Definitions and Conventions}

We use the conventions of \cite{CdRE03} for the metric,
$\g$-matrices and fermions. We freely raise and lower spacetime
indices. No confusion should arise, since contractions are always
performed using the Minkowski metric.

An explicit expression for the tensor $t_8$ is given for example
in \cite{Schwarz}. $t_{abcdefgh}$ is antisymmetric in the pairs
$(ab)$, $(cd)$, etc., and is symmetric under the exchange of such
pairs. It satisfies the following identity:
\begin{eqnarray}
  t_{abcdefgh}M_1^{ab}M_2^{cd}M_3^{ef}M_4^{gh}
  &=&
  -2\big( \tr M_1M_2\,\tr M_3M_4 + \tr M_1M_3\,\tr M_2M_4 + \tr
  M_1M_4\,\tr M_2M_3 \big) \nn\\
  && +8\big( \tr M_1M_2M_3M_4 + \tr M_1M_3M_2M_4 + \tr
  M_1M_3M_4M_2 \big),
\end{eqnarray}
where the $M_i$ are antisymmetric tensors.

The effective action is by definition the generator of 1PI
diagrams:
\begin{equation}
  S_{\rm eff}[A_a] \equiv
  \sum_n\frac{1}{n!}\int\dd^{10}x_1\cdots\dd^{10}x_n\,
  \G^{(n)}_{a_1\cdots a_n}(x_1,\ldots,x_n) A^{a_1}(x_1)\cdots A^{a_n}(x_n),
\end{equation}
hence
\begin{equation}
  \G^{(n)}_{a_1\cdots a_n}(x_1,\ldots,x_n)=
  \left. \frac{\d^n S_{\rm eff}[A_a]}{\d A^{a_1}(x_1)\ldots\d A^{a_n}(x_n)}\right|_{A_a=0}.
\end{equation}
We define the momentum space amplitudes as follows:
\begin{equation}
  (2\pi)^{10}\d^{(10)}(k_1+\ldots+k_n)\G^{(n)}_{a_1\cdots a_n}(k_1,\ldots,k_n)
  \equiv \int\prod_{i=1}^n \dd^{10}x_i\, {\rm e}^{ik_i\cdot x_i}\,
  \G^{(n)}_{a_1\cdots a_n}(x_1,\ldots,x_n).
\end{equation}
An $n$-photon interaction gives the following contribution to the
$S$-matrix:
\begin{equation}
  \mathcal{A}(1,\ldots,n)=i(2\pi)^{10}\delta^{(10)}(k_1+\ldots+k_n)\,
  \z^1_{a_1}\cdots\z^n_{a_n}\G^{(n)}_{a_1\cdots
  a_n}(k_1,\ldots,k_n).
\end{equation}


\section{Proof}

In order to reproduce (\ref{ampl}), we have to obtain the
following 1PI four-point function from (\ref{effactboson}):
\begin{equation}
  \G^{(4)}_{klmn}(k_1,k_2,k_3,k_4) =
  -16(g\a')^2\,t_{akblcmdn}k_1^ak_2^bk_3^ck_4^d\,\mathcal{G}(k_1,k_2,k_3,k_4).
\end{equation}
First we calculate the four-point function in position space:
\begin{multline}
  \G^{(4)}_{klmn}(y_1,\ldots,y_4) =
  \left.\frac{\d^4 S_{\rm eff}[A_a]}{\d A^k(y_1)A^l(y_2)A^m(y_3)A^n(y_4)}\right|_{A_a=0} \\
  = -4!2^4\frac{1}{24}(g\a')^2\int\dd^{10}x
  \left\{\prod_i\dd^{10}x_i\,\d(x-x_i)\right\}D(\pd_{x_1},\ldots,\pd_{x_4})t_{akblcmdn}\\
  \times
  \pd_{x_1}^a\d(x_1-y_1)\pd_{x_2}^b\d(x_2-y_2)\pd_{x_3}^c\d(x_3-y_3)\pd_{x_4}^d\d(x_4-y_4).
\end{multline}
The factor of $2^4$ arises from substituting $F_{ab}=\pd_a A_b
-\pd_b A_a$, the factor 4! from the distributive property of the
functional derivative. To arrive at the result we renamed dummy
variables and made use of the fact that $D$ is symmetric in its
arguments.\\
In momentum space this becomes: {\allowdisplaybreaks
\begin{multline}
  -\frac{1}{16(g\a')^2}(2\pi)^{10}\d(k_1+k_2+k_3+k_4)\G^{(4)}_{klmn}(k_1,k_2,k_3,k_4)
  \\
  \begin{split}
    &=
    \,t_{akblcmdn}\int\dd^{10}x\left\{\prod_i\dd^{10}x_i\dd^{10}y_i\,\d(x_i-x){\rm
    e}^{ik_i\cdot y_i}\right\}D(\pd_{x_1},\ldots,\pd_{x_4}) \\
    &\qquad\times
    \pd_{x_1}^a\d(x_1-y_1)\pd_{x_2}^b\d(x_2-y_2)\pd_{x_3}^c\d(x_3-y_3)\pd_{x_4}^d\d(x_4-y_4) \\
    &=
    \,t_{akblcmdn}\int\dd^{10}x
    \left\{\prod_i \dd^{10}x_i\,\d(x_i-x)\right\}\\
    &\qquad\times
    D(\pd_{x_1},\ldots,\pd_{x_4})\,\pd_{x_1}^a\pd_{x_2}^b\pd_{x_3}^c\pd_{x_4}^d
    \left\{\prod_j\int\dd^{10}y_j\,{\rm e}^{ik_j\cdot
    y_j}\d(x_j-y_j)\right\}\\
    &=
    \,t_{akblcmdn}\int\dd^{10}x
    \left\{\prod_i \dd^{10}x_i\,\d(x_i-x)\right\}\\
    &\qquad\times
    \mathcal{G}(-i\pd_{x_1},\ldots,-i\pd_{x_4})\pd_{x_1}^a\pd_{x_2}^b\pd_{x_3}^c\pd_{x_4}^d
    \left\{\prod_j{\rm e}^{ik_j\cdot x_j}\right\}\\
    &=
    \,t_{akblcmdn}\int\dd^{10}x
    \left\{\prod_i \dd^{10}x_i\,\d(x_i-x)\,{\rm e}^{ik_i\cdot x_i}\right\}
    \mathcal{G}(k_1,\ldots,k_4)\,k_1^ak_2^bk_3^ck_4^d\\
    &=
    \,t_{akblcmdn}\,\mathcal{G}(k_1,\ldots,k_4)\,k_1^ak_2^bk_3^ck_4^d
    \times(2\pi)^{10}\d(k_1+k_2+k_3+k_4).
  \end{split}
\end{multline}}
This completes the proof.



\end{document}